\documentclass[preprint,aps]{revtex4}

\usepackage{graphicx}

\def\eeq{\end{equation}}
\def\beq{\begin{equation}}
\def\bea{\begin{eqnarray}}
\def\eea{\end{eqnarray}}
\begin{document}

\title{Sensitivity to initial conditions in coherent noise models}

\author{Ebru Ergun}
\author{Ugur Tirnakli}%
\email{tirnakli@sci.ege.edu.tr}
\affiliation{%
Department of Physics, Faculty of Science, 
Ege University, 35100 Izmir, Turkey}

\date{\today}

\begin{abstract}
Sensitivity to initial conditions in the coherent noise model of biological 
evolution, introduced by Newman, is studied by making use of damage spreading 
technique. A power-law behavior has been observed, the associated exponent 
$\alpha$ and the dynamical exponent $z$ are calculated. Using these values 
a clear data collapse has been obtained. 

\noindent
{\it PACS Number(s): 89.75.Da, 05.90.+m, 91.30.-f}
\end{abstract}


\maketitle


It is clear that there is an increasing interest in extended 
dynamical systems exhibiting avalanches of activity, whose size 
distribution is scale-free. Some examples of such systems can be 
enumerated as earthquakes \cite{earth}, rice piles \cite{sand}, extinction 
in biology \cite{ext}, evolving complex networks \cite{nets}, etc. 
However, there is no unique and unified theory for such systems. 
One of the possible candidates is the notion of self-organize 
criticality (SOC) \cite{soc}. In SOC models, the whole system is under the 
influence of a small driving force that acts locally and these systems evolve 
towards a critical stationary state having no characteristic spatio-temporal 
scales.  
On the other hand, it is known from literature \cite{newman1} that another kind 
of simple and robust mechanism is available in order for producing scale-free 
behavior. This mechanism is based on the notion of external stress coherently 
imposed on all agents of the system. Since the model does not contain any direct 
interaction among agents, it does not exhibit criticality. Nevertheless, it yields a 
power-law distribution of event size. 
These so-called coherent noise models have been firstly introduced to 
describe large-scale events in evolution, but then they were used as a model of 
earthquakes \cite{newman2} and its aftershock properties \cite{wilkeD} as well as 
aging phenomenon in the model \cite{ugur} have been analysed.

The coherent noise models can be defined as follows: 
Firstly we consider a system, which has $N$ agents. Each agent $i$ has a 
threshold $x_i$ against external stress $\eta$. The threshold levels are 
chosen randomly from some probability distribution $p_{thresh}(x)$. 
The external stress is also drawn randomly from another distribution 
$p_{stress}(\eta)$. An agent is eliminated if it is subjected to the 
stress $\eta$ exceeding the threshold for this agent. 
Algorithmically, dynamics of the model can be given by three steps: 
(i)~at each time step, generate a random stress $\eta$ from 
$p_{stress}(\eta)$, eliminate all agents with $x_i\le \eta$  
and replace them by new agents with new thresholds taken from 
$p_{thresh}(x)$, 
(ii)~select a small fraction $f$ of the $N$ agents at random and 
assign them new thresholds, (iii)~go back to (i) for the next time 
step. 
It is worth mentioning that step (ii) corresponds to the probability for the 
$f$ fraction of the whole agents of undergoing spontaneous transition, which is 
a step necessary for preventing the model from grinding to a halt \cite{wilkeD}.

In the present work, we focus on the sensitivity to the initial conditions 
of the coherent noise models using damage spreading technique. This technique can 
be thought as a method which is borrowed from dynamical systems theory in the 
following sense: 
If we consider two copies of the same one-dimensional dynamical system starting 
from slightly different initial conditions and follow their time evolution, 
we can define the sensitivity function 

\beq
\xi (t) \equiv\lim_{\Delta x(0)\rightarrow 0} 
\frac{\Delta x(t)}{\Delta x(0)} 
=  e^{\lambda t}
\eeq
to quantify the effect of initial conditions. Here, $\Delta x(0)$ and $\Delta x(t)$ 
are the distances between two copies at $t=0$ and $t$ respectively, and $\lambda$ is 
the Lyapunov exponent. If $\lambda > 0$ ($\lambda < 0$), the system is said to be 
{\it strongly sensitive} ({\it strongly insensitive}) to the initial conditions. 
For the marginal case, where $\lambda = 0$, the form of the sensitivity function 
could be a whole class of functions. For the low-dimensional discrete dynamical 
systems, this form is found to be a power-law

\beq
\xi(t) \sim t^{\alpha} \;\;\; .
\eeq 
The $\alpha > 0$ and $\alpha < 0$ cases correspond to {\it weakly sensitive} and 
{\it weakly insensitive} to the initial conditions \cite{maps}.
For the high-dimensional dynamical systems, like the Bak-Sneppen model or like the 
one that we discuss in this work, the same analysis could be performed using 
the Hamming distance instead of the sensitivity function. As in the case of 
low-dimensional systems, one can classify the sensitivity by looking at the 
behavior of the Hamming distance 

\beq
D(t) = \frac{1}{N} \sum_{i=1}^N \left| x_i^{(1)}(t) - x_i^{(2)}(t)\right| \;\; , 
\eeq
where $x_i^{(1)}$ and $x_i^{(2)}$ are two slightly different copies of the 
system under consideration. In this way, up to now, various variants of the 
Bak-Sneppen model have been analysed and related exponents are 
calculated \cite{tamarit,vega1,vega2,ugurlyra1,ugurlyra2}. 
Our aim here is to make the same analysis for the coherent noise models to 
numerically obtain the dynamical exponents of the model.

To proceed further let us define the procedure. We start the simulations from 
a uniform threshold distribution, which means that the system starts from an 
initial event of infinite size that spans the whole system. This is considered 
to be the first copy of the system (namely, $x_i^{(1)}$). In all simulations, 
we use the exponential distribution for the external stress

\beq
p_{stress}(\eta) = a^{-1} \exp\left(-\frac{\eta}{a}\right) \;\;\;\; 
(a>0)\;\; ,
\eeq

\noindent
and the uniform distribution $p_{thresh}(x)$ $(0\le x \le 1)$ for the 
threshold level. 
The second copy (namely, $x_i^{(2)}$) is generated by exchanging the values of 
two randomly chosen sites. Then the dynamics of the model, as explained above, 
is applied for both copies of the system using always the same set of random 
numbers to update them. 
To investigate the properties of sensitivity to initial conditions of the model, 
we follow the temporal evolution of the Hamming distance given in Eq.(3). 
In all cases, we use a large number of realizations to reduce the fluctuations and 
the given results, namely $\left<D(t)\right>$, are the ensemble averages over 
these realizations. From Fig.~1 it is easily seen that the damage spreads as a 
power-law, indicating a weak sensitivity to initial conditions. Obviously, 
the power law growth $\left<D(t)\right>\sim t^{\alpha}$ with 
$\alpha= 0.95 \pm 0.04$ is followed by a plateau with a constant value, 
starting at a certain time depending on the system size. This saturation is 
due to the fact that by this measure one cannot identify that both copies have 
converged to the same random sequence.

\begin{figure}
\includegraphics{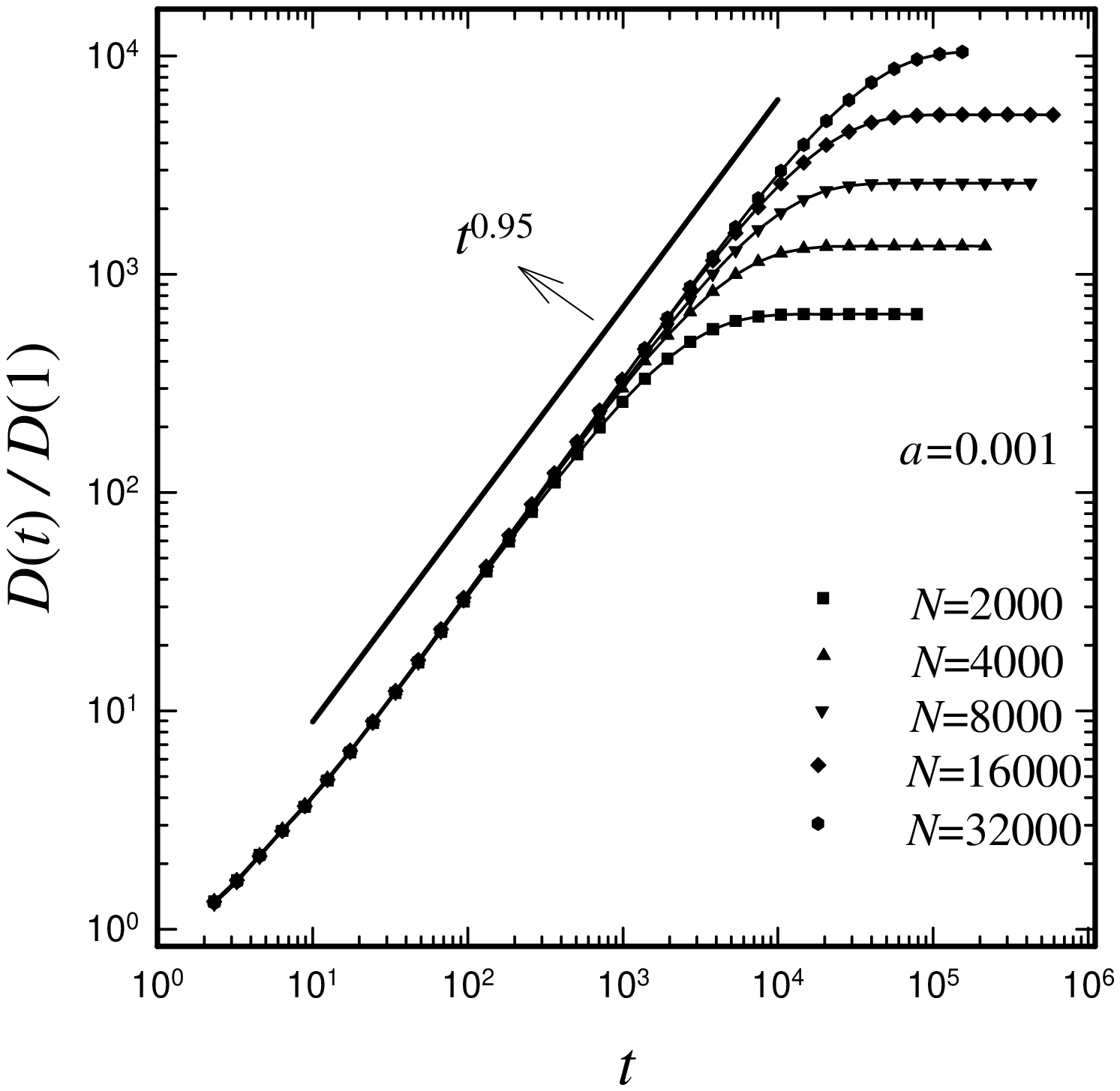}
\caption{\label{fig:Fig1} 
Time evolution of the normalized Hamming distance for five different system 
sizes. For the largest one ($N=32000$) the exponent of the power-law growth 
is estimated as $\alpha=0.95\pm 0.04$ The number of realizations used for 
averaging is 200 for $N=32000$ and 400 otherwise. For a better visualization 
logaritmic binning is employed to all data sets.}
\end{figure}

The second important exponent, called as the dynamical exponent $z$, comes 
from the scaling of $\tau(N)$, which is defined to be the value of $t$ where 
the power-law increasing part crosses over onto the saturation regime. More 
precisely, $\tau$ is the value of $t$ at the intersection point of two 
straight lines, one of which comes from the power-law curve and the other 
from the constant plateau. Then, the scaling $\tau(N)\sim N^z$ defines 
the dynamical exponent $z$, which is obtained from Fig.~2 as 
$z= 0.96 \pm 0.02$.

\begin{figure}
\includegraphics{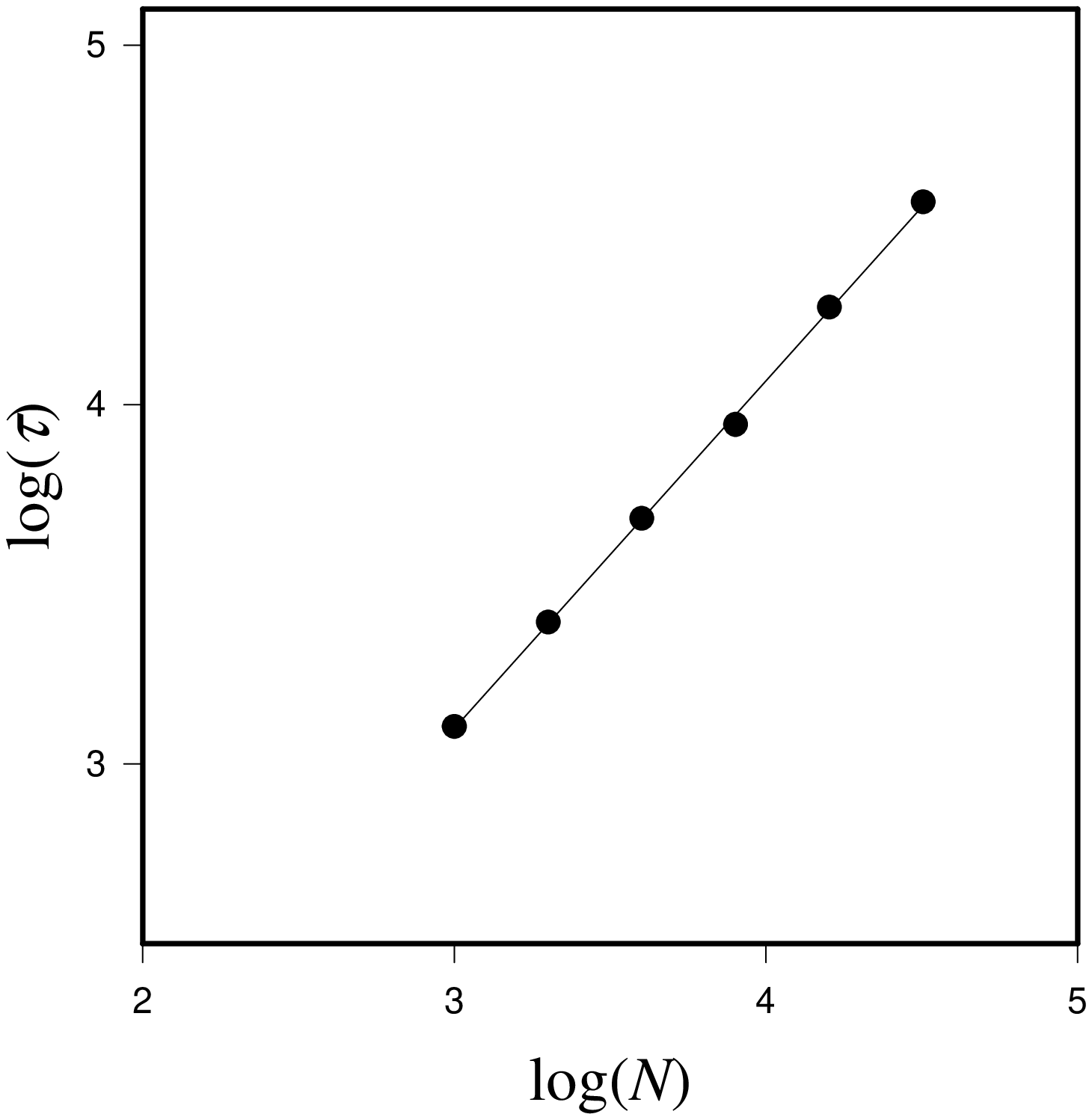}
\caption{\label{fig:Fig2} 
Double log plot of $\tau$ versus $N$ for each curve given in Fig.~1. From this 
scaling we estimate the dynamical exponent as $z=0.96\pm 0.02$.}
\end{figure}

As a final step, we analyse the finite size scaling behavior of the model. 
To accomplish this task, we define the normalized Hamming distance as 

\beq
D(N,t) = \frac{\left<D(t)\right>} {\left<D(1)\right>}
\eeq
and numerically verify that it obeys the scaling ansatz 

\beq
D(N,t) = N^{\beta} F\left(\frac{t}{N^{\gamma}} \right) \;\; ,
\eeq
with $\beta=0.91$, which comes from $\beta=\gamma \alpha$ and 
$\gamma=z$. A clear data collapse can easily be seen in Fig.~3.  
It is worth noting that the results presented here do not depend on 
the choice of the numerical value of the model parameter $a$ 
and/or the fraction $f$. In our simulations we use $a=0.001$ and 
different $f$ values for each system size in order to assure only one 
agent is eliminated in step (ii) to reduce the computational time.

\begin{figure}
\includegraphics{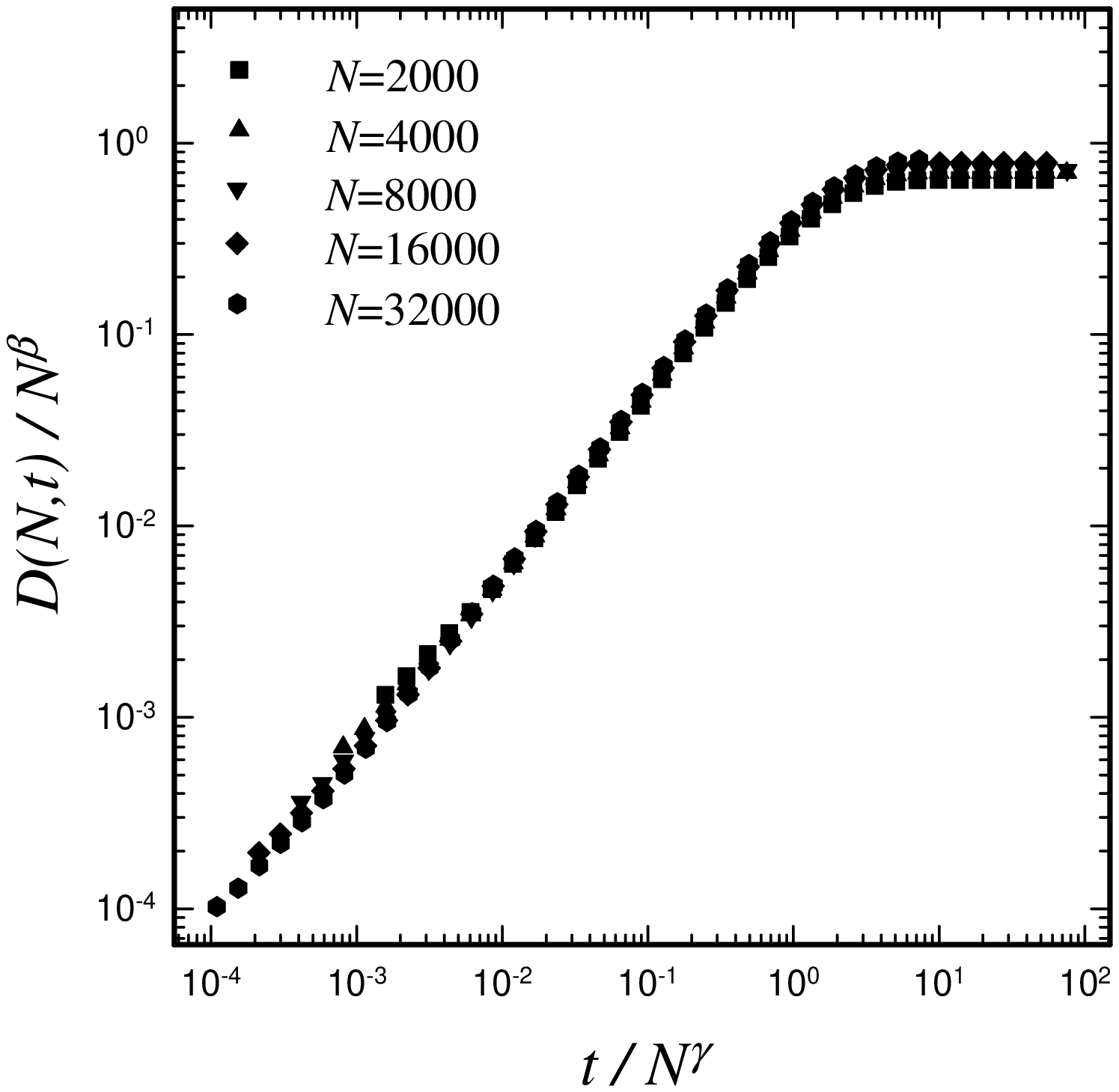}
\caption{\label{fig:Fig3} 
Data collapse of the finite size scaling given in Eq.(6) for the same 
system sizes used in Fig.~1 and Fig.~2.}
\end{figure}

Summing up, we have studied the sensitivity to initial condition properties 
of the coherent noise models using damage spreading technique. 
We found that the model exhibits weak sensitivity to initial conditions, 
a property which is common for other high-dimensional dynamical systems 
like Bak-Sneppen model and its variants as well as one- and two-dimensional 
discrete systems like logistic map families. 
The numerically obtained values of the power-law exponent $\alpha$ and 
the dynamical exponent $z$ are appeared to be different from those of 
the Bak-Sneppen model, which signals out the discrepancy between the 
universality classes of these models.

\section*{Acknowledgments}
This work is supported by the Turkish Academy of Sciences, 
in the framework of the Young Scientist Award Program 
(UT/TUBA-GEBIP/2001-2-20). 

\end{document}